\documentclass[usenatbib]{mn2e}
\usepackage{epsfig}
\usepackage{amssymb}
\usepackage{lscape,graphicx}
\usepackage{longtable}
\usepackage{psfig}
\usepackage{lscape}
\usepackage{rotating}
\usepackage{times}
\usepackage{graphicx}
\usepackage{epstopdf}

\def\gsim{\mathrel{\raise0.35ex\hbox{$\scriptstyle >$}\kern-0.6em
\lower0.40ex\hbox{{$\scriptstyle \sim$}}}}
\def\lsim{\mathrel{\raise0.35ex\hbox{$\scriptstyle <$}\kern-0.6em
\lower0.40ex\hbox{{$\scriptstyle \sim$}}}}

\title[{\it Herschel} observations of \hbox{[OI]}63\,$\mu$m in
$z\sim1$ SMGs]{{\it
    Herschel}-PACS observations of \hbox{[OI]}63\,$\mu$m towards
  submillimetre galaxies at $z\sim$1$^\star$}

\author[Coppin et al.]{
\parbox[t]{\textwidth}{
K.E.K. Coppin$^{1}$, A.L.R. Danielson$^{2}$, J.E. Geach$^{1}$, J.A. Hodge$^{3}$,
A.M. Swinbank$^{2}$, J.L. Wardlow$^{4}$, F. Bertoldi$^{5}$, A. Biggs$^{6}$,
W.N. Brandt$^{7,8}$, P. Caselli$^{9}$, S.C. Chapman$^{10}$,
H. Dannerbauer$^{11}$, J.S. Dunlop$^{12}$, T.R. Greve$^{13}$, F. Hamann$^{14}$, R.J. Ivison$^{12,15}$,
A. Karim$^{2}$, K.K. Knudsen$^{16}$, K.M. Menten$^{5}$,
E. Schinnerer$^{3}$, Ian Smail$^{2}$, M. Spaans$^{17}$,
F. Walter$^{3}$, T.M.A. Webb$^{1}$, P.P. van der Werf$^{18}$}\\\\
$^{1}$Department of Physics, McGill University, 3600 Rue University,
Montr\'eal, QC, H3A 2T8, Canada\\
$^{2}$ Institute for Computational Cosmology, Durham University, South
Road, Durham DH1 3LE\\
$^{3}$ Max-Planck-Institut f\"{u}r Astronomie, K\"{o}nigstuhl 17,
Heidelberg, D-69117, Germany\\
$^{4}$ Department of Physics and Astronomy, University of California,
4129 Fredrick Reines Hall, Irvine, CA 92697-4575, USA\\
$^{5}$ Max-Planck-Institut f\"{u}r Radioastronomie, Auf dem H\"{u}gel 69, Bonn, D-53121, Germany\\
$^{6}$European Southern Observatory, Karl-Schwarzschild Stra\ss e 2,
D-85748 Garching, Germany\\
$^{7}$ Department of Astronomy and Astrophysics, Pennsylvania State
University, 525 Davey Lab, University Park, PA 16802, USA\\
$^{8}$Institute for Gravitation and the Cosmos, Pennsylvania State University, University Park, PA 16802, USA\\
$^{9}$ School of Physics and Astronomy, University of Leeds, Leeds LS2 9JT\\
$^{10}$ Institute of Astronomy, University of Cambridge, Madingley Road, Cambridge CB3 0HA\\
$^{11}$ Universit\"{a}t Wien, Institut f\"{u}r Astrophysik,
T\"{u}rkenschanzstra\ss e 17, 1180 Wien, Austria\\
$^{12}$ SUPA, Institute for Astronomy, University of Edinburgh, Royal Observatory, Blackford Hill, Edinburgh EH9 3HJ\\
$^{13}$ Department of Physics \& Astronomy, University College London,
Gower Street, London WC1E 6BT\\
$^{14}$Department of Astronomy, University of Florida, Gainesville, FL 32611-2055, USA\\
$^{15}$ UK Astronomy Technology Centre, Royal Observatory, Blackford Hill, Edinburgh EH9 3HJ\\
$^{16}$ Department of Earth and Space Sciences, Chalmers University of Technology, Onsala Space Observatory, SE-43992 Onsala, Sweden\\
$^{17}$ Kapteyn Astronomical Institute, University of Groningen, PO Box 800, 9700 AV, Groningen, The Netherlands\\
$^{18}$ Leiden Observatory, Leiden University, PO Box 9513, NL-2300 RA
Leiden, the Netherlands
}

\begin{document}

\maketitle

\begin{abstract}
We present {\it Herschel}-PACS spectroscopy of the [OI]63\,$\mu$m
far-infrared cooling line from a sample of six unlensed and 
spectroscopically-confirmed 870\,$\mu$m-selected submillimetre (submm)
galaxies (SMGs) at $1.1<z<1.6$ from the LABOCA Extended Chandra Deep
Field South (ECDFS) Submm Survey (LESS). 
This is the first survey of [OI]63\,$\mu$m, one of the main photodissociation region (PDR)
cooling lines, in SMGs. New high-resolution
ALMA interferometric 870\,$\mu$m continuum imaging confirms
that these six {\it Herschel}-targeted SMG counterparts are
bona fide sources of submm emission.  We detect [OI]63\,$\mu$m in
two SMGs with a SNR$\gsim3$, tentatively detect [OI]63\,$\mu$m in one
SMG, and constrain the line flux for
the non-detections. We also exploit the
combination of submm continuum photometry from 250--870\,$\mu$m and our
new PACS continuum measurements to constrain the far-infrared
luminosity, L$_{\rm FIR}$, in these SMGs to $\lesssim30$\,per cent.  
We find that SMGs do not show a deficit
in their [OI]63\,$\mu$m--to--far-infrared (FIR) continuum luminosity
ratios (with ratios ranging from $\simeq0.5$--1.5\,per cent), similar
to what was seen previously for the [CII]158\,$\mu$m--to--FIR ratios
in SMGs.  These observed ratios are about an order of magnitude higher than what is
seen typically for local ultra luminous infrared galaxies (ULIRGs), 
which adds to the growing body of evidence that SMGs are not simply
`scaled up' versions of local ULIRGs.  Rather, the PDR
line--to--L$_{\rm FIR}$ ratios suggest that the star formation
modes of SMGs are likely more akin to that of local normal (lower-luminosity)
star-forming galaxies, with the bulk of the star formation occurring in
extended galaxy-scale ($\sim$kpc) regions.  These observations
represent the first step towards a census of the major PDR cooling
lines in typical SMGs that will be attainable with ALMA, 
enabling detailed modelling to probe the global properties of the star
formation and the evolutionary status of SMGs.

\end{abstract}


\noindent $^\star${\small {\it Herschel} is an ESA space
    observatory with science instruments provided by European-led
    Principal Investigator consortia and with important participation
    from NASA.}

\section{Introduction}

Far-infrared bright galaxies comprise some of
the most luminous galaxies in the local Universe, but they contribute
just 0.3 per cent of the total local galaxy luminosity density \citep{SandMir96}. At
first glance it appears that they are an unimportant component of the
galactic zoo, however, sensitive (sub)millimetre cameras have
revealed that ultraluminous infrared galaxies (ULIRGs; L$_{\rm FIR}\geq10^{12}$\,L$_\odot$) are much more
abundant at high redshifts ($z\gtrsim2$). Since their discovery
(\citealt{Smail97}; \citealt{Hughes98}; \citealt{Barger98}), nearly 13 years of piece-meal survey work has so far
produced samples totalling $\sim500$ 850--1200-selected (sub)mm galaxies (SMGs;
e.g.~\citealt{Coppin06}; \citealt{Austermann10}; \citealt{Weiss09}),
with extreme star formation rates (SFRs) of
$\sim100$--1000\,M$_\odot$\,yr$^{-1}$ (e.g.~\citealt{Kovacs06};
\citealt{Coppin08}; \citealt{Magnelli12}) and fuelled by vast reservoirs of cold molecular gas 
(M$_\mathrm{H_{2}}\sim5\times10^{10}$\,M$_\odot$; \citealt{Greve05};
\citealt{Tacconi06,Tacconi08}; \citealt{Bothwell12}).  Spectroscopic follow-up of a subset
of SMGs suggests that the volume density of ULIRGs increases
$\simeq1000$ times out to $z\sim2$ (\citealt{Chapman05};
\citealt{Banerji11}). Therefore, luminous obscured galaxies
likely dominate the total bolometric emission from star formation at
early epochs  (e.g.~\citealt{Pascale09}; \citealt{Marsden09};
\citealt{Eales09,Eales10}; \citealt{Oliver10}), and possibly
correspond to the active starburst progenitor phase of massive spheroidal
galaxies (e.g.~\citealt{Lilly99}).

However, several critical questions about SMGs remain unanswered.  The
most fundamental of these is: \textit{How is the immense luminosity
  of an SMG powered? -- by an intense single compact nuclear starburst (as seen in
  local ULIRGs) or by a more extended
  starburst (as seen in local normal star-forming galaxies)?}  One route to tackle this question is to ask what are the
local analogues of SMGs and can these be used to understand the
physics of the star formation activity within the distant SMG
population?  Are SMGs simply `scaled up' ULIRGs, where the
star formation occurs in a compact and highly obscured nuclear region -- 
potentially with a significant contribution from an active galactic
nucleus (AGN; e.g.~\citealt{Alexander05b})? Or does some fraction of
the population resemble very scaled up `normal' star-forming infrared
(IR) galaxies, which have far-IR emission arising from an extended,
cooler component (termed `cirrus'; \citealt{ERR03})?  
If we could empirically match subsets of the SMG population to
local galaxies, then this would provide significant insights into
the physical processes occuring within this important population.
The synergy of \textit{Herschel} and the Atacama
Large Millimetre Array (ALMA) provides a novel
opportunity to move beyond the first generation of detection
experiments by beginning to examine the interplay between the
interstellar medium (ISM) and the power source of high-redshift luminous dusty star-forming galaxies.  
\smallskip

\subsection{Far-infrared spectroscopy: a probe of SMG astrophysics}

The primary diagnostic lines of the ISM in the far-IR have traditionally included the
  fine-structure lines of [OIII]52, [NIII]57, [OI]63, [OIII]88,
  [NII]122, [OI]146, [CII]158\,$\mu$m, [CI]370\,$\mu$m, and [CI]610\,$\mu$m.  The relative strengths of these lines can be modeled as a function of the neutral gas cloud density and the
 cloud-illuminating far-UV radiation field intensity in photodissociation
 regions (PDRs; e.g.~\citealt{Tielens85}; \citealt{Kaufman99};
 \citealt{Wolfire03}; \citealt{Meijerink05}), providing diagnostics of the physical
conditions of the ISM. While much has been learned about the ISM
conditions in Galactic star-forming regions and local
galaxies through facilities such as the {\it Infrared Space Observatory
  (ISO}; e.g.~\citealt{Malhotra01}; \citealt{Luhman98,Luhman03};
\citealt{Negishi01}) and the Kuiper Airborne
Observatory (KAO; e.g.~\citealt{Crawford85}; \citealt{Stacey91}; \citealt{Poglitsch95}), due to previously poor instrument sensitivity and
inaccessibility from the ground, dedicated
surveys probing the ISM conditions in samples of high redshift dusty
galaxies have only recently begun in earnest (e.g.~\citealt{Stacey10};
\citealt{Decarli12})
with the commissioning of
purpose-built instrumentation like the redshift ($z$)
and Early Universe Spectrometer (ZEUS; \citealt{Stacey07};
\citealt{HD09}).  For galactic and extragalactic systems, the
`classical' PDR tracers are the brightest fine-structure emission
lines [CII]158\,$\mu$m and [OI]63\,$\mu$m, acting as
the primary coolants for the 
dense ($n\gtrsim10$--10$^{5}$\,cm$^{-3}$), warm (T$\sim100$--1000\,K),
 and neutral material.  Since [CII]158\,$\mu$m and [OI]63$\mu$m
 correlate well with SFR and are sufficiently
 luminous (accounting for as much as $\sim0.1$--1\% of the
 bolometric luminosities in nearby galaxies and are thus much more
 luminous than molecular lines; \citealt{Stacey91}), they have the potential to be used as
 powerful tracers of ISM conditions out to high-$z$.  However, \textit{ISO} observations of local
 ULIRGs revealed that the [CII]158\,$\mu$m emission is 
about an order of magnitude lower relative to the far-IR (FIR) continuum flux
 than is seen in normal and starburst galaxies
 (e.g.~\citealt{Malhotra01}; \citealt{Luhman03}).  And this has
 recently been shown by \citet{JGC11} using {\it Herschel}-PACS to be a general aspect of far-IR fine
 structure lines (see Section~\ref{discuss} and \citealt{JGC11} for a discussion of a possible physical
 explanation). The potential weakness of [CII]158\,$\mu$m has
 implications for its use as a star formation tracer
 for high-redshift studies ($z>4$;
 e.g.~\citealt{Maiolino05}; although cf.~\citealt{Maiolino09}; \citealt{DeBreuck11}).
 In order to make further progress in understanding the
driver of star formation in luminous high-redshift SMGs in relation to
galaxy populations in the local Universe, we need to begin compiling a
similar set of (multi-transition) CO, [CII]158\,$\mu$m,
and [OI]63\,$\mu$m data (which provide strong joint constraints on the PDR and star
formation conditions such as the Hydrogen volume density, $n$, and the
far-UV radiation field strength, $G_{0}$ in Habing units) for typical luminous star-forming SMGs at the epoch where their population peaks ($z\sim$1--3).  

\subsection{This paper: New {\it Herschel} observations of \hbox{[OI]} 63\,$\mu$m in SMGs}

In the present paper we will be discussing observations performed with
the ESA {\it Herschel Space Observatory} \citep{Pilbratt10}, in
  particular employing {\it Herschel's} large telescope and powerful
  science payload to perform photometry and spectroscopy using the Spectral
  and Photometric Imaging REceiver (SPIRE; \citealt{Griffin10}) and
  the Photodetector Array Camera and Spectrometer (PACS; \citealt{Poglitsch10}).
This paper presents the first attempt at compiling a
census of the strength of [OI]63\,$\mu$m in far-IR luminous SMGs at
the peak of star formation activity in the Universe ($z\sim$1--2),
enabling us to determine its suitability as a star formation
tracer at high redshift.  These \textit{Herschel} observations
represent a key stepping stone to future ALMA studies of
[CII]158\,$\mu$m and (multi-transition) CO to investigate the ISM physics
of SMGs in detail, since [OI]63\,$\mu$m emission can only be
measured in $1\lesssim z\lesssim 2$ SMGs with {\it Herschel}-PACS, as
the emission line falls outside of the ALMA bands for all but the
highest redshifts ($z\geq4$).  

This paper is organized as follows: the sample selection, {\it
  Herschel}-PACS spectroscopic observations and data reduction are
described in Section~\ref{obsdr}. 
In Section~\ref{analysis}, we present
the main analysis and results of the PACS spectrocopy and combine these
measurements with L$_{\rm FIR}$ estimates obtained from the combination of
Large Apex Bolometer Camera (LABOCA; \citealt{Siringo09}) and new {\it
  Herschel}-SPIRE and PACS photometry.  We discuss the implications
of our results in Section~\ref{discuss}. Finally, our conclusions are given in
Section~\ref{conclusions}.  Throughout the paper we assume
cosmological parameters of $\Omega_\Lambda=0.73$, $\Omega_\mathrm{m}=0.27$,
and $H_\mathrm{0}=71$\,km\,s$^{-1}$\,Mpc$^{-1}$ \citep{Spergel03}.

\section{Observations and data reduction}\label{obsdr}

%
%Table 1
%
\begin{table*}
\begin{minipage}{1.0\textwidth}
%\scriptsize
\caption{Observation details and measured [OI]63\,$\mu$m integrated
  line fluxes and continuum flux densities from our {\it
    Herschel}-PACS programme (OT1\_kcoppin\_1).  
SMG optical counterpart positions from \citet{Wardlow11} are given in
columns 3--4, 
to which the zLESS and archival spectroscopy are referenced.  The
optical spectroscopic redshifts in column 5 are from: (a)
zLESS (Danielson et al., in preparation); (b) \citet{Vanzella05}; and (c)
\citet{Silverman10}. Columns 6--9 give the
Herschel Observation Date (OD), Observation identification number
(ObsID), the number of range repetitions per nod cycle times the
number of nod cycles, and the total Astronomical Observation Request (AOR) duration (including the slew).
Column 10 gives the measured (continuum-subtracted) integrated
[OI]63\,$\mu$m flux and error from the Gaussian fit described in
Section~\ref{spectra}, or the 3-$\sigma$ upper limit assuming a
Gaussian profile with a FWHM$=300$\,km\,s$^{-1}$ and a peak height of $3\times\sigma_{\rm rms}$.  
Column 11 gives the best-fitting continuum flux density (or upper
limit) at the spectrum centre (indicated in square brackets).  Column
12 gives the 1\,$\sigma$ rms of the entire spectrum, quoted for the representative
bin width at the centre of each band (see square brackets in previous column).  We note
that the line and continuum flux density error budgets do not include the estimated additional
absolute calibration rms uncertainty of 12 per cent (see text).  }\label{tab:obs}
\scriptsize
\hspace{-0.2in}
\begin{tabular}{llrrrrrccccc}
\hline
\multicolumn{1}{c}{SMG IAU name} & \multicolumn{1}{c}{Nickname} & \multicolumn{3}{c}{\underline{Counterpart Parameters}}
& \multicolumn{4}{c}{\underline{Observation Parameters}} &
\multicolumn{2}{c}{\underline{Spectral Measurements}}\\
& & \multicolumn{1}{c}{R.A.} & \multicolumn{1}{c}{Dec.} &
\multicolumn{1}{c}{z$_\mathrm{opt}$} &
\multicolumn{1}{c}{OD} & \multicolumn{1}{c}{ObsID} &
\multicolumn{1}{c}{$n_\mathrm{rep} \times n_\mathrm{cyc}^\mathrm{a}$}
& \multicolumn{1}{c}{AOR} & \multicolumn{1}{c}{[OI]63\,$\mu$m
  Flux} & \multicolumn{1}{c}{Continuum Flux} &
\multicolumn{1}{c}{Spectrum rms}\\
& & \multicolumn{2}{c}{(J2000)} & & & & & \multicolumn{1}{c}{(s)} &
    \multicolumn{1}{c}{($\times10^{-18}$\,W\,m$^{-2}$)} &
    \multicolumn{1}{c}{(mJy)} & \multicolumn{1}{c}{($\times10^{-19}$\,W\,m$^{-2}$)}\\
\hline
LESS\,J033329.9--273441 & LESS21 & 3:33:29.73 & -27:34:44.4 & 1.235$^{a}$ & 1020 & 1342239701
& $4\times2$ & 8146 & $<3.8$  & $<18$ [141\,$\mu$m] & 3.9 \\
LESS\,J033217.6--275230 & LESS34 & 3:32:17.61 & -27:52:28.1 & 1.0979$^{b}$ & 1020 & 1342239703
& $4\times2$ &
7874 & $<3.9$ & $<17$ [133\,$\mu$m] & 4.1 \\
LESS\,J033331.7--275406 & LESS66 & 3:33:31.92 & -27:54:10.3 & 1.315$^{c}$ & 1013 & 1342239369
& $4\times2$ &
8349 & $2.0\pm0.6$ & $27\pm7$ [146\,$\mu$m] & 4.1\\
LESS\,J033155.2--275345 & LESS88 & 3:31:54.81 & -27:53:41.0 & 1.269$^{a}$ & 1020 & 1342239705
& $4\times2$ &
8214 & $1.8\pm0.6$ & $23\pm7$ [143\,$\mu$m] & 4.2\\
LESS\,J033140.1--275631 & LESS106 & 3:31:40.17 & -27:56:22.4 & 1.617$^{a}$ & 1019 &
1342239753 & $5\times2$ &
11444 & $3.8\pm1.4$ & $<23$ [165\,$\mu$m] & 4.4\\
LESS\,J033150.8--274438 & LESS114 & 3:31:51.09 & -27:44:37.0 & 1.606$^{a}$ & 1020 &
1342239702 & $5\times2$ &
11444 & $<5.6$ & $48\pm11$ [165\,$\mu$m] & 5.3\\
\hline
\end{tabular}
\normalsize
\end{minipage}
\end{table*}

\subsection{Sample selection and multi-wavelength properties}\label{sample}

The requirement of our survey is a sufficient number of SMGs selected
in a uniform manner that have been spectroscopically confirmed to lie
between $0.7<z<2$ so that the redshifted [OI]63$\mu$m emission line
will lie within the most sensitive region of the first order of the PACS
spectrometer (107--189\,$\mu$m; \citealt{Poglitsch10}).  The SMGs
also need to be accessible by ALMA for future studies
(i.e.~$\delta\lesssim30^{\circ}$).  One such field is that
covered by the 870\,$\mu$m 
LABOCA Extended Chandra Deep Field South (ECDFS) Submillimetre Survey 
(LESS; \citealt{Weiss09}), which comprises a parent sample of
126 SMGs detected above 3.7-$\sigma$ in the largest contiguous
($30'\times30'$) and uniform map with a 1\,$\sigma$ rms of
$\sim1.2$\,mJy.  \citet{Biggs11} have identified secure radio and
mid-IR counterparts for 63\,per cent of these SMGs.  Since the ECDFS is one of
the pre-eminent cosmological deep fields and has as such been extensively
imaged and targeted spectroscopically, some of
these counterparts already possess secure archival redshifts.
Redshifts for the remaining counterparts without archival spectroscopy
were targeted through a large observational programme (zLESS; Danielson et  al., in
preparation) using the European Southern Observatories (ESO) Very Large
Telescope (VLT) FOcal Reducer and low dispersion Spectrograph v.2
(FORS2) and VIsible MultiObject Spectrograph (VIMOS).  
At the time the {\it Herschel} proposal was submitted in 2010, the VLT survey was
$\approx40$\,per cent complete.  The total number of SMGs fulfilling the
requirements outlined above at that time was nine, and these nine SMGs
thus comprise our {\it Herschel}-PACS spectroscopy sample.  Thus, our
sample is complete in terms that it is comprised of all of the
spectroscopically confirmed $1<z<2$ SMGs in LESS that are radio and/or
24\,$\mu$m identified.

We make use of additional multi-wavelength data sets available for the
radio and/or 24\,$\mu$m LESS SMG counterparts in order to further interpret
our results, which we now briefly describe.  \citet{Wardlow11} 
measured the optical-mid-IR photometry for the robustly identified LESS SMG counterparts and
provide photometric redshift estimates and L$_{\rm IR}$ estimates.  In
order to assess any potential AGN contribution in the SMGs, we
use the {\it Spitzer} InfraRed Array Camera (IRAC;
\citealt{Fazio04}) photometry from \citet{Wardlow11} extracted from the 
{\it Spitzer} IRAC/MUSYC Public Legacy Survey in the ECDFS
(SIMPLE; \citealt{Damen11}) and the 250\,ks and 4\,Ms X-ray imaging
and catalogues of the ECDFS/CDFS (\citealt{Lehmer05}; \citealt{Xue11}).
Here we also make use of new public {\it Herschel} Multi-tiered
Extragalactic Survey 
(HerMES; \citealt{Hermes}) {\it Herschel}-SPIRE maps
to extract 250--500\,$\mu$m photometry in combination 
with the zLESS spectroscopic redshifts, providing the tightest
constraints on L$_{\rm FIR}$ currently possible for these SMGs
(to $\lesssim30$\,per cent).  In addition, high-resolution 
($\sim1.55''$ full width at half maximum; FWHM) ALMA 870\,$\mu$m continuum data have been
collected for the majority of the 126 LESS SMGs during cycle 0 
to a 1\,$\sigma$ rms depth of $\sim$0.4\,mJy, yielding positions accurate to
$<0.3''$ (Karim et al., in preparation). The ALMA data have enabled us to verify unambiguously the validity of the radio and/or mid-IR
counterparts for our SMG sample, which are used here as a basis for the PACS target
positions and central frequency tuning of the spectrometer (see
Section~\ref{summary} for a summary of the ALMA ID results for our sub-sample).

\subsection{{\it Herschel}-PACS spectroscopy observations}\label{obs}

The [OI]63\,$\mu$m emission line in our sample of SMGs was observed with the \textit{Herschel Space
Observatory}'s PACS red camera over 112-167\,$\mu$m using
the high-resolution range scan spectroscopy mode and employing a chop/nod to provide
background subtraction.  For each target, we covered a small spectral range
of $\approx4$\,$\mu$m, noting that only the central $\approx3$\,$\mu$m of the
spectra are usable, corresponding to $\Delta z \simeq \pm 0.03$--0.05
to account for small redshift errors and/or systematic offsets from the optical
redshift, as well as to allow for a reliable continuum or baseline fit.
The aim was to reach a uniform 1\,$\sigma$ line sensitivity of
$\simeq0.8\times10^{-18}$\,W\,m$^{-2}$ across the sample (see
Table~ref{tab:obs} for the actual sensitivities that were reached).
Six targets were observed in 2012 February over a total of
15.4\,hrs (including overheads).  Table~\ref{tab:obs} gives the target list, observation mode parameters, and
individual integration times.  We note that an additional three sources
were targeted as part of this programme (LESS10, LESS50 and LESS118;
see \citealt{Biggs10}), but the IDs (and therefore redshifts) have been
subsequently revised based on new ALMA continuum data (Karim et al., in preparation; see also Section~\ref{summary}).  Thus, these
three observations do not provide interesting constraints on the
[OI]63\,$\mu$m emission from SMGs and so we have dropped them from
the analysis.

\subsection{Data reduction}\label{dr}

\begin{figure*}
\epsfig{file=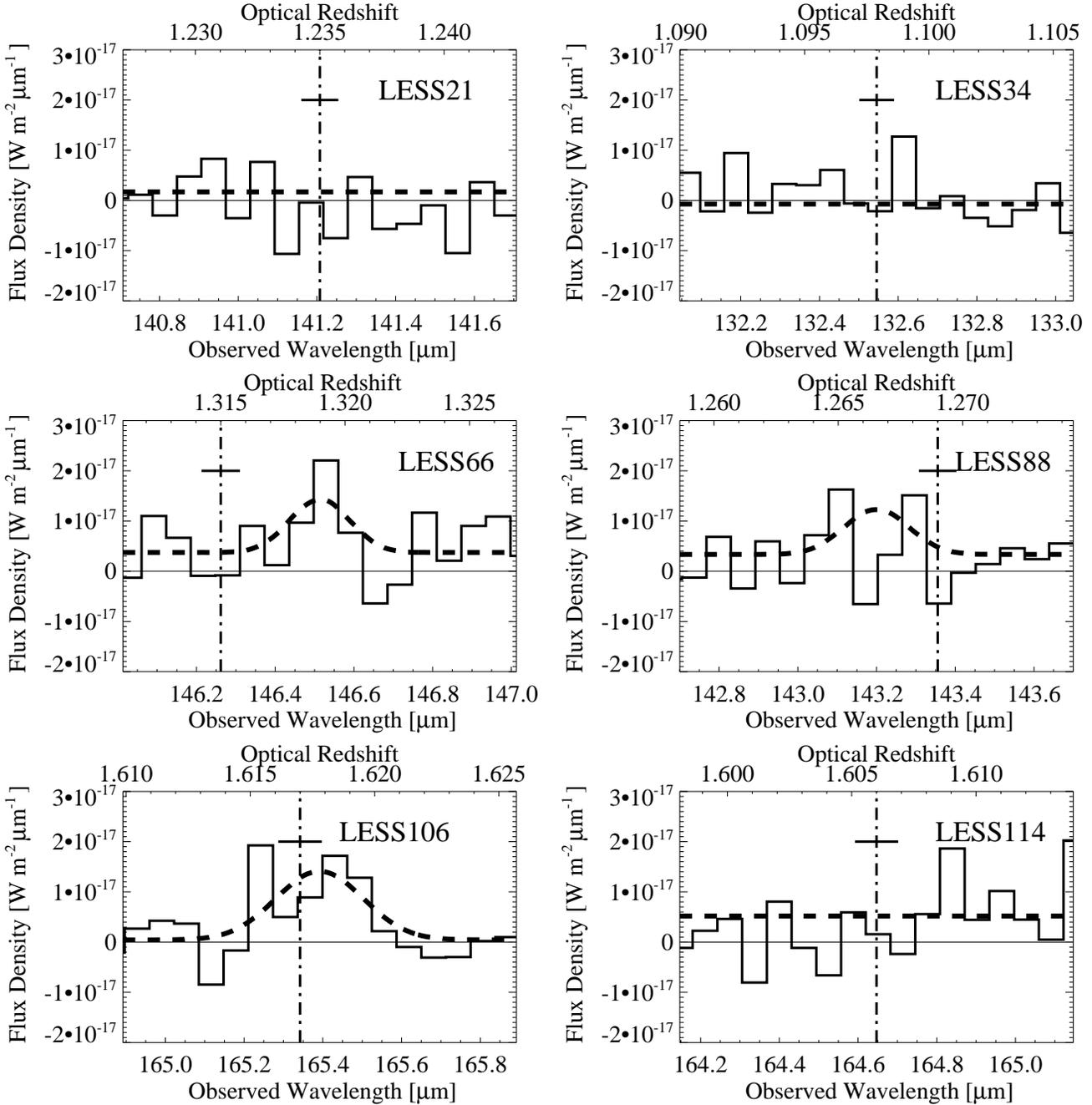,width=1\textwidth}
\caption{Observed infrared {\it Herschel}-PACS Nyquist sampled spectra of our sample of six
  SMGs.  The spectra have been zoomed to a 1\,$\mu$m window centred on
  the optical or [OI]63\,$\mu$m derived redshift, and 
the dot-dashed vertical line and
  error bar indicate the expected position of
  the [OI]63\,$\mu$m emission line based on prior optical spectroscopy and an
  approximate 1\,$\sigma$ error of 100\,km\,s$^{-1}$. For LESS66,
  LESS106 (formal detections) and LESS88 
(only marginally detected) the best-fitting Gaussian+continuum profile
is overplotted as a dashed curve (see
  Section~\ref{spectra} for details). For the non-detections (LESS21, LESS34,
  and LESS114) the best
  underlying continuum fit is overplotted as a dashed line.}
\label{fig:spectra}
\end{figure*}

The data were reduced with the \textit{Herschel} interactive processing
environment ({\sc hipe} v8.1.0; \citealt{Ott10}). We employed the
background normalisation pipeline script distributed with {\sc hipe}, which implements an alternative way to correct for drifts in the response of PACS during the observation and an alternate flux calibration than the standard pipeline. Using this alternative pipeline script results in marginally better
quality data products with smoother continua, which is helpful
when searching for faint spectral features atop a faint continuum.   As part of the standard pipeline, the continuum in each of the 16 spectral pixels was scaled
to the median value to correct for residual flat field effects, and
finally the two nod positions were combined to completely remove the
sky (telescope) background. Given that our targets are point sources
to PACS we have measured line fluxes from spectra extracted from the central
$9.4''\times9.4''$ spatial pixel of the $5\times5$\,pixel
field-of-view of the PACS spectrometer.  
The pipeline automatically applies a beam size correction
factor and an additional in-flight correction of the absolute
response.  The spectra have been re-binned to Nyquist sample the spectral
resolution element, which is appropriate for resolved lines (see
Fig.~\ref{fig:spectra}).  We have chosen not to
upsample\footnote{Upsampling refers to increasing the sampling rate of
a signal.  Here we use upsample=1.  If upsample$>1$ is used, then 
the frequency bins are not independent (ie.~they will have some data
in common) and this will have the effect of changing the spectral
noise properties as well as the line widths.} the data,
so that each bin is independent, yielding reliable
estimates of the noise properties in the spectra. A
similar data reduction strategy was adopted by the Survey with
Herschel of the ISM in Nearby Infrared Galaxies (SHINING) team for reducing faint spectra of
high-redshift galaxies (priv. comm. E. Sturm and J. Graci\'{a}-Carpio).

Third order data in the
wavelength range 51--73\,$\mu$m are collected by the blue camera in parallel with the first order [OI]63\,$\mu$m data.  However, these observations are not frequency-tunable and cover
only $\sim2$\,$\mu$m. Nevertheless, the data were reduced using the same procedure as above
and examined for any serendipitous line detections, but the useful
parts of the spectra do not cover any useful known lines at our source redshifts and
so we do not discuss these data further.

\section{Analysis and results}\label{analysis}

We present new [OI]63\,$\mu$m data for the SMGs, including two new
detections of the emission line, and one marginal detection. 
For the non-detections, upper limits are only
useful if {\it both} the counterpart and redshift are correct: this {\it
  Herschel} spectroscopy program was designed to target only the most statistically probable `robust' counterparts
also posessing good quality redshifts.  However, given the significant time lag from the survey
definition to the time of data acquisition ($\approx1.5$ years), new data
have since been accumulated from zLESS and ALMA that can
help us to verify if the targeted counterparts and redshifts
originally chosen are still correct. In the following section we
examine and discuss each PACS spectrum.  We then derive the FIR continuum flux and L$_{\rm FIR}$ for each SMG
via a modified blackbody spectral energy distribution (SED) fit to publicly
available submm data so that we can
explore the [OI]63\,$\mu$m--to--FIR continuum ratios of our sample as
a function of L$_{\rm FIR}$ in context with other low- and
high-redshift galaxy populations.

\subsection{Spectral line and continuum measurements}\label{spectra}

We detect weak [OI]63\,$\mu$m emission in LESS66, LESS106 and
tentatively in LESS88 (see Fig.~\ref{fig:spectra}).  We can derive useful upper
limits for the cases where the line emission is not detected, and only in those cases where we are
certain that the correct redshift and/or SMG counterpart
was targeted. We detect continuum emission in
LESS66, LESS88, and LESS114 at $\gsim3\,\sigma$, and for the continuum
non-detections we derive $3\,\sigma$ upper limits to the continuum flux density in the PACS
spectral range.

For the detections, a Gaussian profile and continuum are fitted
simultaneously to the Nyquist sampled spectra by performing non-linear Levenberg-Marquardt
least-squares minimization using the {\sc MPFIT IDL} package
\citep{Markwardt08}. Since the pipeline-calculated errors in the PACS spectrum are known to be unreliable, 
we have clipped off the noisy edges from each spectrum
($\approx1$\,$\mu$m on each side) and weight each of the remaining
central channel data points equally.  To calculate the
velocity-integrated flux of the [OI]63\,$\mu$m emission line, we integrate the best-fitting (continuum-subtracted) Gaussian
profile. The 1\,$\sigma$ errors on the velocity-integrated flux and
continuum flux density are derived from a bootstrap estimate generated by adding random noise with a 1\,$\sigma$ rms
equivalent to the 1\,$\sigma$ rms of the line-free channels, repeated
1000 times with replacement (e.g.~see \citealt{Wall03}). This should
provide a reasonable estimate of the uncertainties since our choice of
spectral binning (upsample=1) ensures that the noise should not be
strongly correlated between channels. 

For the non-detections we calculate 3\,$\sigma$ upper limits assuming
a Gaussian profile with a 3\,$\sigma_{\rm rms}$ peak height centred at the 
expected position of the [OI]63$\mu$m emission line.  We use the
channel-to-channel 1\,$\sigma$ rms noise across the clipped spectra and
adopt an instrumental resolution-corrected emission line FWHM of 300\,km\,s$^{-1}$, which is consistent with our detections
and those of other high-redshift galaxies that have been observed with PACS (e.g.~\citealt{Sturm10}).  

Integrated emission line fluxes, upper limits and continuum flux densities are summarized
in Table~\ref{tab:obs}.  Note that the rms uncertainties associated with the absolute calibration of PACS have not been
folded into the quoted flux uncertainties and are estimated to be $\simeq12$ per cent\footnote{See http://herschel.esac.esa.int/twiki/bin/view/Public/PacsCalibrationWeb}.

In the following we first summarize the counterpart identifications in light of the new
ALMA data, and then discuss the details of the counterpart identifications and redshift
robustness for each targeted SMG in turn, along with the new PACS spectroscopy
results.  Reference to a `robust' counterpart is meant in a statistical
sense: given a potential radio or mid-IR counterpart with its flux density and
radius from the SMG position, the a priori probability, $p$, of finding at
least one object within that radius of at least that flux density from
the expected number of events at random is $p\leq0.05$.
A `tentative' counterpart refers to sources with $0.05<p\leq0.1$ and
is often still low enough to indicate a likely counterpart.  \citet{Ivison02,Ivison07} and \citet{Biggs11} provide a more in depth
explanation on this subject.  Even though this is the only feasible route for
precisely pin-pointing the positions of the bulk of SMGs in the
absence of very high-resolution interferometric continuum 
submm data (ie.~in the pre-ALMA era), this is only a statistical identification
method.  Recent submm inteferometric observations of SMGs
have demonstrated that potentially a large fraction of `robust' SMG identifications may be more complicated (with
up to $\sim30$\,per cent breaking up into multiple components; see
e.g.~\citealt{Cowie09}; \citealt{Wang11}; \citealt{Hatsukade10}).
Armed with new deep high-resolution ALMA 870\,$\mu$m continuum data
(Karim et al., in preparation), we can test
the reliability of the counterpart identifications using the radio or
mid-IR, which is used here as a basis for targeting the
[OI]63\,$\mu$m emission line with {\it Herschel}-PACS.  Redshifts for
the counterparts were obtained by an ESO VLT FORS2/VIMOS Large Programme (zLESS; Danielson et al., in
preparation) and are supplemented by the wealth of existing public
archival spectroscopy in the ECDFS.  The quality of the redshifts are
categorized by zLESS as follows: Q=1 refers to a secure
redshift (e.g.~based on the identification of multiple lines) and Q=2 refers to a probable
redshift (e.g.~based on a single bright emission line identification).

\subsubsection{Summary of the counterpart ID re-examination in light
  of new ALMA data}\label{summary}

In summary, after examining the preliminary new ALMA continuum data from
Karim et al.\ (in preparation), it appears that our
original sample of nine targets (see Section~\ref{sample}) should be
reduced to six.  The ALMA data indicate that we have not
targeted a counterpart associated with submm
continuum emission in two cases (LESS10 and LESS118) and we
thus discard these sources from our sample and from further
discussion.  The ALMA data have also not been able to validate the ID
for LESS50 (whether this is because the submm emission is resolved out
in the ALMA map or that the SMG has split up into several fainter
sources when seen by ALMA -- none of which correspond to the ID), 
and so to be conservative we also discard this source from
the sample.  Although, we note that our discarded targets are still radio or
24\,$\mu$m emitting galaxies (just not the source of the submm
emission) with good quality redshifts and thus remain useful data
for other purposes.  
The six remaining targets are associated with ALMA continuum emission
and possess a good quality redshift so we can safely measure line
fluxes or sensitive upper limits in these cases: LESS21, LESS34,
LESS66, LESS88, LESS106, and LESS114.

In a few cases (LESS34, LESS106, LESS114) it appears that the
LABOCA-detected SMG may be splitting up into two or more continuum
sources when seen by ALMA at much higher resolution. 
For these, ideally the LABOCA and SPIRE photometry should first be
deblended or deconvolved before calculating the L$_{\rm FIR}$ values
in Section~\ref{sed}.  Unfortunately,
without complete redshift information and multiple interferometric
photometry for all of the ALMA continuum
sources this is impossible to do in a robust way, and it is beyond
the scope of this paper.  The diagnostic we are interested in comparing with literature
values is the [OI]63\,$\mu$m-to-FIR continuum ratio versus L$_{\rm FIR}$ for the
SMGs.  If the FIR continuum is reduced accordingly, the
[OI]63\,$\mu$m--to--FIR continuum ratio will increase, and so 
this has little bearing on our main finding in this paper that this
ratio is high for SMGs compared to local galaxies with similar L$_{\rm
  FIR}$ (see Section~\ref{ratio}).  Thus, by not correcting the FIR
continuum luminosity we follow a conservative approach.

\subsubsection{LESS21}

It has a tentative 24\,$\mu$m counterpart (e.g.~borderline secure, with
$p=0.067$), and there is ALMA continuum emission coincident with this
position, confirming the ID.

The zLESS spectroscopic redshift of this counterpart is $z=1.2350$.  
The zLESS VIMOS spectrum of this counterpart is consistent with the photometric redshift
solution of $z=1.26^{+0.074}_{-0.183}$ from \citet{Wardlow11}, with
the tentative detection of MgII absorption.  It is not detected in the
X-ray, and has a 3\,$\sigma$ upper limit on its
full-band flux of S$_{0.5-8\,{\rm keV}}<6.1\times10^{-16}$\,erg\,cm$^{-2}$\,s$^{-1}$, derived from
the 250\,ks X-ray imaging of the ECDFS \citep{Lehmer05}.    
The S$_{8\mu{\rm m}}$/S$_{4.5\mu{\rm m}}$ colour ratio is $\simeq0.7$, indicating that
the galaxy contains a negligible contribution to its energetics from
an AGN (see \citealt{Coppin10}). 

Given the optical redshift, the [OI]63\,$\mu$m emission line is expected to lie at
141.21\,$\mu$m, however no line emission near this wavelength is
apparent in the spectrum. We thus calculate a 3\,$\sigma$ upper limit to the
emission line flux.

\subsubsection{LESS34}

It neighbours LESS10 ($\approx 20$\,arcsec away) and has a single secure 24\,$\mu$m counterpart (and
formally no secure radio counterpart), which was
targeted with PACS.  There does appear to be some weak ALMA continuum emission coincident
with this targeted ID, however there is also another bright ALMA
source lying just outside of the SMG counterpart radius adopted by
\citet{Biggs11}.  The archival redshift of our targeted source, $z=1.0979$ (Q=1), is from \citet{Vanzella05}.  The single SMG identified with LABOCA is thus possibly comprised of multiple
sources of submm emission, which would have the effect of a reduced
inferred L$_{\rm FIR}$ for our targeted counterpart.  It
is not detected in the X-ray, and has a 3\,$\sigma$ upper limit to its
full-band flux of S$_{0.5-8\,{\rm
    keV}}<7.0\times10^{-17}$\,erg\,cm$^{-2}$\,s$^{-1}$, derived from
the 4\,Ms X-ray imaging of the CDFS \citep{Xue11}.  
The S$_{8\mu{\rm m}}$/S$_{4.5\mu{\rm m}}$ colour ratio is $\simeq0.2$, indicating that
the galaxy contains a negligible contribution to its energetics from
an AGN (see \citealt{Coppin10}). 

Given the redshift, we expect the
[OI]63\,$\mu$m emission line to lie at 132.55\,$\mu$m.  We do not
see an emission line in the spectrum near this wavelength and so a 3\,$\sigma$ upper limit to the
emission line flux is calculated.

\subsubsection{LESS66}

It has a single robust radio and 24\,$\mu$m counterpart.  The
ALMA continuum data reveal 
continuum emission coincident with this ID. The zLESS redshift of
this counterpart from an excellent quality
(Q=1) VIMOS spectrum is indicated to be $z=1.310$, although zLESS FORS2
spectroscopy indicates $z=1.314$ (Q=1).  A secure (Q=1) Keck-based
spectroscopic redshift of $z=1.315$ was obtained by
\citet{Silverman10} from a program to target optical and near-IR
counterparts of faint X-ray sources in the ECDFS.  Interestingly,
\citet{Silverman10} classify this source as a Broad-Line AGN (BLAGN) based on
the presence of at least one emission line having a
FWHM$>2000$\,km\,s$^{-1}$.  It is detected in the X-ray with a
full-band flux of S$_{0.5-8\,{\rm keV}}=(31.6\pm1.4)
\times10^{-15}$\,erg\,cm$^{-2}$\,s$^{-1}$ \citep{Lehmer05}.  
Its X-ray flux is consistent with that expected for an optically classified
QSO given its submm flux density and is higher than typically seen in 
SMGs ($\sim10^{-16}-10^{-14}$\,erg\,cm$^{-2}$\,s$^{-1}$; \citealt{Alexander05b}). 
Although there is clearly a luminous AGN in this
SMG, its S$_{8\mu{\rm m}}$/S$_{4.5\mu{\rm m}}$ colour ratio of $\simeq1.1$
indicates that the AGN is likely not dominating its mid--to--far-IR
energetics (see \citealt{Coppin10}). 

Given the optical redshifts, the [OI]63\,$\mu$m line is expected to lie between
145.95--146.20\,$\mu$m.  We find a weak $\simeq3$\,$\sigma$ detection of
[OI]63 centred at $146.51\pm0.07\,\mu$m (corresponding to a redshift of $z=1.319$), 
offset by $\simeq500$\,km/s from the Keck optical redshift ($z=1.315$), which is 
similar in magnitude to the difference between the two zLESS optical
redshifts.  This shift in the [OI]63\,$\mu$m
emission from the optical-based redshift is likely a real effect since
offsets of this magnitude are often seen in BLAGN. 
The best-fitting FWHM of the line emission is $178\pm113$\,km\,s$^{-1}$, which is
narrower than the instrumental resolution at this wavelength
($\simeq256$\,km\,s$^{-1}$), which is unphysical. We thus proceed to
fit the spectrum holding the line FWHM fixed to the instrumental
resolution to yield a less biased line flux (although we note that
this will artificially decrease the quoted error budget on the fitted
integrated line flux in Table~\ref{tab:obs}).  In any case, the
best-fitting observed line width (the line is unresolved) 
implies that there is no sign of an outflow with
velocities significantly above the typical Keplerian motions of an SMG
($\sim$ a few 100 km\,s$^{-1}$).  This despite the high SFR of LESS66 of
$\simeq370$\,M$_{\odot}$\,yr$^{-1}$ (see Table~\ref{tab:ir}) which would
yield $\approx1$ supernova (SN) per year assuming a \citet{Salpeter55}
initial mass function (IMF), resulting in a significant input of mechanical
energy (1\,SN$\sim10^{44}$\,J) into the ISM.

\subsubsection{LESS88}

It has a single secure robust radio counterpart, which is also detected at 24\,$\mu$m (formally
a tentative counterpart).  The ALMA data show a single continuum source coincident with this ID. The
zLESS spectroscopic redshift is $z=1.269$ (Q=2).   It
is not detected in the X-ray, and has a 3\,$\sigma$ upper limit to its
full-band flux of S$_{0.5-8\,{\rm
    keV}}<4.1\times10^{-16}$\,erg\,cm$^{-2}$\,s$^{-1}$, derived from
the 250\,ks X-ray imaging of the ECDFS \citep{Lehmer05}. Its S$_{8\mu{\rm
m}}$/S$_{4.5\mu{\rm m}}$ colour ratio is $\simeq0.8$, indicating that
an AGN is unlikely to be contributing significantly to the energetics of the
system (see \citealt{Coppin10}). 

Given the optical redshift, the [OI]63\,$\mu$m line is expected to lie at
143.36\,$\mu$m, and there appears to be some weak emission near this position in
the spectrum and we claim a tentative detection of [OI]63\,$\mu$m in
LESS88. The emission is weak, which precludes a successful
Gaussian fit.  We thus fix the centre and height of a Gaussian by eye
and proceed to fit for the FWHM (resulting in a FWHM of $307\pm277$\,km\,s$^{-1}$
which has been corrected for an instrumental resolution of
261\,km\,s$^{-1}$ at this wavelength).  We note that holding the two
Gaussian parameters fixed will artificially decrease the quoted error budget on the fitted
integrated line flux in Table~\ref{tab:obs}.  This results in a reasonable
line shape and we overplot it on Fig.~\ref{fig:spectra}.  The line
emission is offset approximately by $-$300\,km\,s$^{-1}$ from the optical
redshift, noting that offsets of $\sim200$--300\,km\,s$^{-1}$
from the H$\alpha$ and CO derived redshifts are often seen in infrared-bright
high-redshift star-forming galaxies (e.g.~\citealt{Sturm10};
\citealt{Steidel10}).

\subsubsection{LESS106}

It has a single
secure robust IRAC 8.5\,$\mu$m counterpart which is also detected in the radio
(formally only a tentative ID on its own) and at 24\,$\mu$m (not a formal robust
or tentative ID).  This
is the ID we targeted with PACS.  This counterpart also shows some 
coincident ALMA continuum emission, although we note that the emission
is rather weak and that the sensitivity of the beam is only
$\sim38$\,per cent at this location.  There are actually a few weak ALMA
continuum sources present in the beam, and so several sources could be responsible for producing the
submm flux density seen as a single SMG in the LABOCA map.   The zLESS spectroscopic redshift of our
target is $z=1.617$ (Q=2).   It is detected in the X-ray with a
full-band flux of S$_{0.5-8\,{\rm keV}}=(3.4\pm0.5)\times10^{-15}$\,erg\,cm$^{-2}$\,s$^{-1}$
\citep{Lehmer05}, which is consistent with typical X-ray detected 
SMGs ($\sim10^{-16}-10^{-14}$\,erg\,cm$^{-2}$\,s$^{-1}$; \citealt{Alexander05b}). 
Although there is an AGN in this
SMG, its S$_{8\mu{\rm m}}$/S$_{4.5\mu{\rm m}}$ colour ratio is $\simeq0.8$, indicating that
an AGN is unlikely to be contributing significantly to the energetics of the
system (see \citealt{Coppin10}). 

Given the optical redshift of our {\it Herschel}-targeted counterpart, $z=1.617$, the [OI]63\,$\mu$m line is expected to lie at
165.34\,$\mu$m, and a $\sim3$\,$\sigma$ emission line is seen at this
wavelength (and consistent with the 1\,$\sigma$ error on the optical
spectroscopic redshift).  The line has a FWHM of
$413\pm254$\,km\,s$^{-1}$ (corrected for an instrumental resolution of
227\,km\,s$^{-1}$ at the observed wavelength).  We note that this FWHM
is similar to the mean CO FWHM for SMGs ($510\pm80$\,km\,s$^{-1}$; \citealt{Bothwell12}).

%
%Table 2
%
\begin{table*}
\begin{minipage}{1.0\textwidth}
\caption{We have listed the submm flux densities and our SED fit constraints
for the six SMGs in our {\it Herschel}-PACS sample.  The 250--500\,$\mu$m flux densities and
instrumental noises are given and have been extracted from the HerMES SPIRE maps at the
SMG counterpart location.  We note that we have added a confusion noise of 4.8, 5.5, and
6.1\,mJy \citep{Nguyen10} in quadrature to the instrumental noise estimates at 250,
350, and 500\,$\mu$m, respectively, for the SED fitting, as per the
standard practice.  The
870\,$\mu$m flux density is the deboosted flux density value from \citet{Weiss09}.
The FIR continuum flux, L$_{\rm FIR}$, T$_{\rm d}$ and SFR are 
derived following the SED fitting procedure outlined in Section~\ref{sed},
holding $\beta=1.5$ and using the FIR and L$_{\rm FIR}$ definitions from
\citet{SandMir96}. The SFR is calculated following
\citet{Kennicutt98}.}\label{tab:ir}
\centerline{
\hspace{-0.2in}
\begin{tabular}{lrrrrcccc}
\hline
\multicolumn{1}{c}{SMG nickname} & \multicolumn{1}{c}{S$_{250}$} &
\multicolumn{1}{c}{S$_{350}$} & \multicolumn{1}{c}{S$_{500}$} &
\multicolumn{1}{c}{S$_{870}$} & \multicolumn{1}{c}{FIR continuum flux} & \multicolumn{1}{c}{L$_{\rm FIR}$} &
\multicolumn{1}{c}{T$_{\rm d}$} &
\multicolumn{1}{c}{SFR}\\
& \multicolumn{1}{c}{(mJy)} & \multicolumn{1}{c}{(mJy)} &
\multicolumn{1}{c}{(mJy)} & \multicolumn{1}{c}{(mJy)} &
\multicolumn{1}{c}{($\times10^{-16}$\,W\,m$^{-2}$)} &
\multicolumn{1}{c}{($\times10^{12}$\,L$_{\odot}$)} & \multicolumn{1}{c}{(K)} &\multicolumn{1}{c}{(M$_{\odot}$\,yr$^{-1}$)} \\
\hline
LESS21 & $26.1\pm3.5$ & $22.7\pm2.9$ & $22.4\pm3.8$ & $7.6\pm1.3$ &
$1.3\pm0.4$ & $0.72\pm0.12$ & $22\pm2$ & $100\pm30$ \\%Lir=0.6+/-0.18 d12
LESS34 & $23.9\pm2.2$ & $31.0\pm2.1$ & $26.8\pm2.6$ & $6.3\pm1.3$ &
$1.4\pm0.4$ & $0.59\pm0.09$ & $21\pm2$ & $80\pm20$ \\%Lir=0.47+/-0.13 d12
LESS66 & $19.7\pm2.2$ & $16.1\pm2.2$ & $2.9\pm2.5$ & $5.3\pm1.5$ &
% Lir=2.1+/-0.6 d12 (helou with dale correction)
$4.1\pm1.2$ & $1.4\pm0.3$ & $39\pm5$ & $370\pm100$\\ % assuming 1.315 - same as 1.318
LESS88 & $15.8\pm2.1$ & $24.2\pm2.1$ & $19.9\pm2.5$ & $4.5\pm1.4$ &
$1.7\pm0.6$ & $0.75\pm0.16$ & $25\pm3$ & $140\pm50$ \\%Lir=0.83+/-0.28 d12
LESS106 & $24.6\pm2.2$ & $24.3\pm2.1$ & $18.1\pm2.5$ & $4.0\pm1.4$ &
$2.3\pm0.8$ & $1.5\pm0.4$ & $30\pm3$ & $350\pm120$ \\%Lir=2.0+/-0.7 d12
LESS114 & $42.9\pm2.2$ & $38.2\pm2.2$ & $34.8\pm2.5$ & $3.9\pm1.4$ &
$5.7\pm1.0$ & $3.3\pm0.5$ & $36\pm3$ & $850\pm150$ \\%Lir=4.9+/-0.9 d12
\hline
\end{tabular}
}
\end{minipage}
\end{table*}

\subsubsection{LESS114}

It has a single secure
robust radio and 24\,$\mu$m counterpart, which was targeted with PACS.  The zLESS FORS2 spectroscopic redshift
is $z=1.606$ (Q=1).   This counterpart also shows some ALMA continuum
emission.  However, there appears to be a nearly-equivalently bright ALMA
continuum source present in the field, and so it is likely that two sources could be responsible for producing the
submm flux density seen as a single SMG in the LABOCA map.   
It is detected in the X-ray with a flux of S$_{0.5-8\,{\rm
    keV}}=(2.0\pm0.5)\times10^{-15}$\,erg\,cm$^{-2}$\,s$^{-1}$
\citep{Lehmer05}, which is consistent with typical X-ray detected 
SMGs ($\sim10^{-16}-10^{-14}$\,erg\,cm$^{-2}$\,s$^{-1}$; \citealt{Alexander05b}). 
Although there is an AGN in this
SMG, its S$_{8\mu{\rm m}}$/S$_{4.5\mu{\rm m}}$ colour ratio is $\simeq0.7$, indicating that
an AGN is unlikely to be contributing significantly to the energetics of the
system (see \citealt{Coppin10}). 

Given the optical redshift, the [OI]63\,$\mu$m line is expected to lie at
164.65\,$\mu$m, but there is no apparent emission near this
position in the spectrum.  We therefore derive a 3\,$\sigma$ upper limit to the line
flux.

\begin{figure*}
\epsfig{file=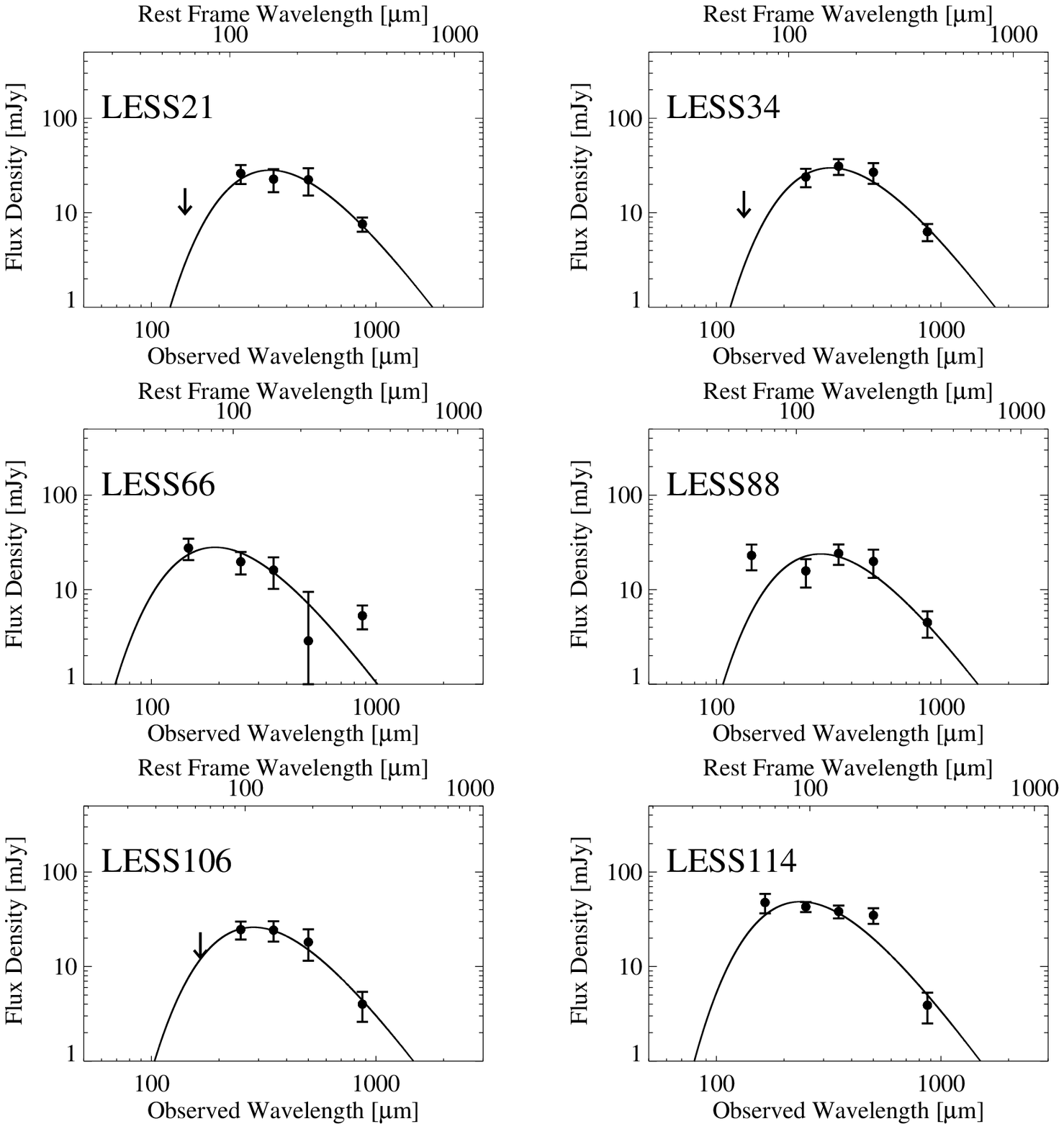,width=1.\textwidth}
\caption{Best-fitting modified single-temperature blackbody SEDs for our sample of
  SMGs fit to 250--870\,$\mu$m photometry (see
  Table~\ref{tab:ir}) as well as our new PACS continuum flux densities
  or upper limits (see Table~\ref{tab:obs}). 
  Here $\beta$ has been fixed to 1.5, and the best-fitting T$_{\rm d}$
  and derived values (the FIR continuum flux and L$_{\rm FIR}$) are
  listed in Table~\ref{tab:ir}. See Section~\ref{sed} for
  further details.}
\label{fig:seds}
\end{figure*}

\subsection{Far-infrared SED fitting}\label{sed}

We have constrained the far-IR SEDs of our SMG sample by fitting their
observed submm photometry with a simple modified blackbody spectrum
of the form $B_{\nu} \propto
\nu^{3}/[\mathrm{exp}(h\,\nu/k\,T_\mathrm{d})-1]$ multiplied by an
emissivity function $\propto \nu^{\beta}$, where the dust
temperature, $T_\mathrm{d}$, determines the location of the SED peak
and $\beta$ is the power law index of our simplified dust emissivity
power law.  Observations of local and distant galaxies constrain
$\beta$ to between 1 and 2 in the Rayleigh-Jeans side of the emission
spectrum so we fix $\beta=1.5$ (e.g.~\citealt{Hildebrand83};
\citealt{Agladze94}; \citealt{Dunne01}).  By fixing $\beta$, we are assuming that the dust
properties are similar in all the SMGs, but note that we do not account for possible systematics
in our quoted errors.  For reference, changing the value of $\beta$ by
$\pm0.5$ changes the best-fitting dust temperatures by about
$\pm5$--10\,K and the integrated IR luminosities by $\pm5$--15\,per cent.
As in Section~\ref{spectra}, we have used the
least-squares minimization {\sc MPFIT IDL} package from
\citet{Markwardt08} to perform the fitting.

The submm photometry used in each SED fit
includes the deboosted LABOCA 870\,$\mu$m flux densities from
\citet{Weiss09}, SPIRE 250, 350, and 500\,$\mu$m photometry (see
Table~\ref{tab:ir}), and
also our new PACS continuum flux densities 
or upper limits (from 114--165\,$\mu$m, depending on the redshifted
[OI]63\,$\mu$m location; see Table~\ref{tab:obs}).  
We have extracted the SPIRE 250, 350, and 500\,$\mu$m flux densities and
instrumental errors for our SMGs from the publicly available HerMES DR1 maps\footnote{The public SPIRE maps used in
  this paper were obtained through the {\it Herschel} Database in Marseille,
  HeDaM (hedam.oamp.fr/HerMES).} \citep{Hermes} at the radio
or 24\,$\mu$m counterpart positions.  Our SPIRE flux density
measurements should be less biased on average than other
methods that could be employed, such as peak matching or measuring the
flux density at the submm position, since there is less combined
uncertainty in the true source position when using the more precise
radio or mid-IR positions
to measure the SPIRE flux densities (e.g.~\citealt{Coppin08}).  
We have added confusion noise estimates of 4.8, 5.5, and 6.1\,mJy
from \citet{Nguyen10} in quadrature to
the instrumental errors at 250, 350 and 500\,$\mu$m, respectively, and these
errors are used in the SED fitting routine, as per the standard
practice (e.g.~\citealt{Magnelli12}).  

\begin{figure*}
\epsfig{file=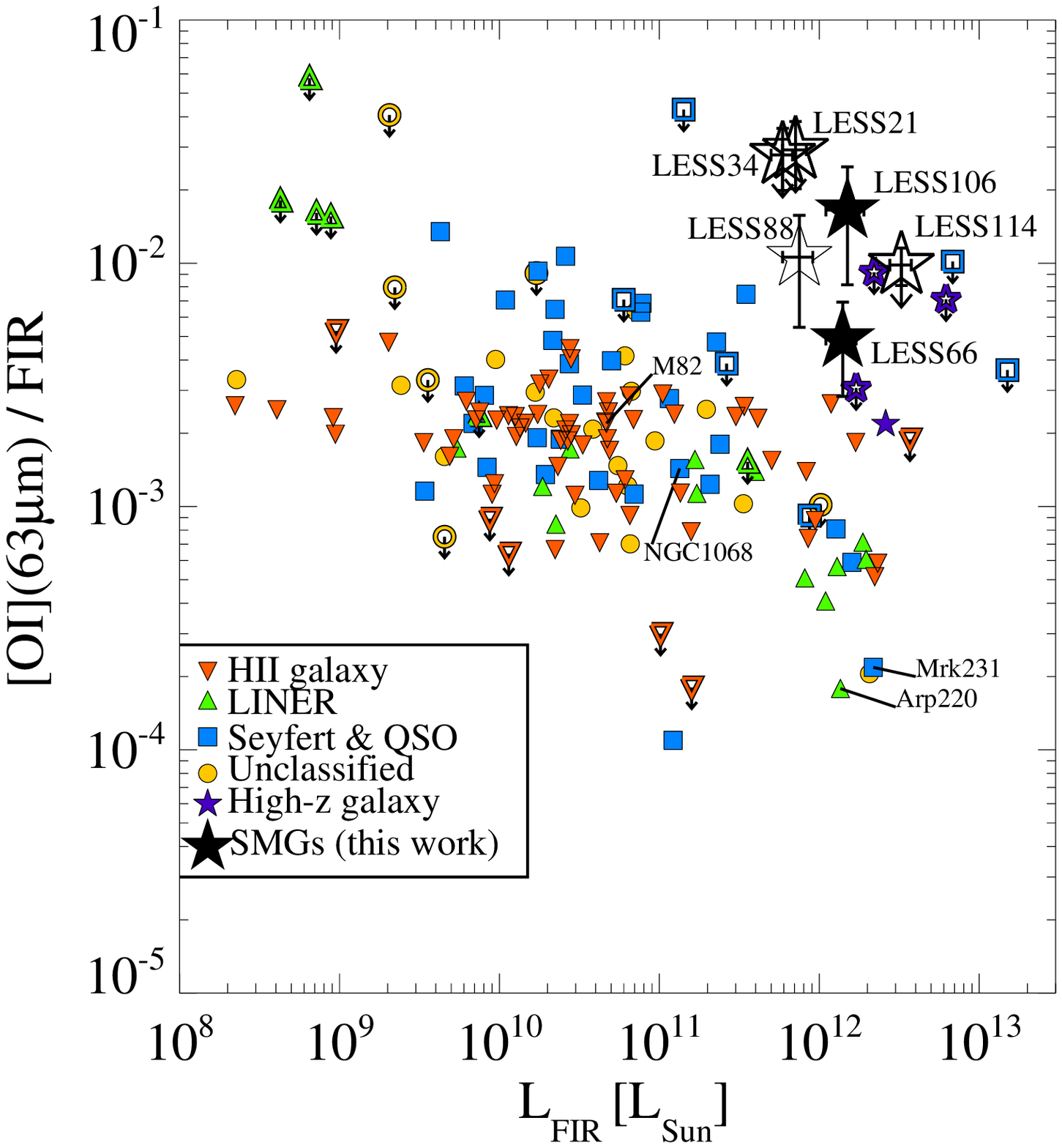,width=1.\textwidth}
\caption{The observed [OI]63\,$\mu$m--to--FIR continuum ratio as a
  function of L$_{\rm FIR}$ for SMGs from this paper (large black stars) compared with available
  literature values (compiled by \citealt{JGC11}). The literature
  compilation includes all galaxies observed with {\it ISO}
  (\citealt{Colbert99}; \citealt{Negishi01}; \citealt{Malhotra01}; \citealt{Luhman03};
  \citealt{Dale04}; \citealt{Brauher08}) and with PACS
  \citep{JGC11}.  The high-redshift ($1.13<z<3.04$) galaxy
  comparison sample includes 3C\,368 \citep{Brauher08}, SMM\,J2135-0102 \citep{Ivison10}, SDP.81
  \citep{Valtchanov11}, and MIPS\,J142824.0+352619 \citep{Sturm10}, all of which have been corrected for
  gravitational lensing. Several familiar well-studied local galaxies are
  indicated on the plot for reference, including M82, NGC1068 (a
  LIRG), and ULIRGs Arp220 and Mrk231 (\citealt{Fischer10};
  \citealt{JGC11}). The solid symbols represent the two SMGs with
  [OI]63\,$\mu$m detections, and the open symbol represents the SMG
  which we have `tentatively' detected.  The open symbols with downward arrows indicate
  3\,$\sigma$ upper limits. In Section~\ref{sed}, we found that the
  FIR continuum flux and L$_{\rm FIR}$ are possibly underestimated by
  estimated factors of 2.8 and 1.7, respectively, for LESS21 and LESS34, and by factors
  of 1.6 and 1.5, respectively, for LESS106.  Applying these correction factors to
  these SMGs would move these sources down and to the right in the plot
by the same factors. The SMGs do not show an [OI]63\,$\mu$m
  emission line deficit compared to local ULIRGs, similar to what is
  seen for the [CII]158\,$\mu$m emission line for SMGs.}
\label{fig:main}
\end{figure*}

We then calculate L$_{\rm FIR}$ and the FIR continuum flux
using the following common definitions for ease of comparing with the literature
compilation of \citet{JGC11} who use the same definitions.  We integrate the best-fitting rest frame SED between
40--500\,$\mu$m to derive L$_{\rm FIR}$, as defined by
\citet{SandMir96}.  Following \citet{SandMir96}, we also calculate the
FIR continuum flux by integrating the rest frame SED between 42.5--122.5\,$\mu$m and dividing this by
$4 \pi$\,D$^{2}_{\rm L}$, where D$_{\rm L}$ is the luminosity
distance.   The errors on FIR and L$_{\rm FIR}$ are derived from the formal
1\,$\sigma$ errors in the fitted parameters computed from the
covariance matrix returned by {\sc MPFIT}.  
Because our single temperature modified
blackbody model does not reproduce the Wien side of the far-IR
SED well for some of the SMGs (see Fig.~\ref{fig:seds}) we are likely to be underestimating the
true value of FIR and L$_{\rm FIR}$ in these cases. To check this we use
the range of SEDs from the \citet{DaleHelou02} template library
normalised such that they are consistent with our measured photometry
in order to estimate approximate correction
factors to our values in Table~\ref{tab:ir} for each SMG (which are of
course dependent on the SED templates used).  We find that our FIR
continuum flux and L$_{\rm FIR}$ values for both LESS21 and LESS34 in Table~\ref{tab:ir} are
potentially underestimated by factors of $\simeq2.8$ and 1.7, respectively, and
that our FIR continuum flux and L$_{\rm FIR}$ values for LESS114 are
potentially underestimated by factors of $\simeq1.6$ and 1.5, respectively.
Applying these correction factors to these SMGs would move these sources down and to the right in Fig.~\ref{fig:main}
by the same factors.  We find negligible correction factors for our other sources, whose SEDs
are better sampled at rest frame wavelengths $\lesssim$120\,$\mu$m
(see Fig.~\ref{fig:seds}). 

We calculate the SFR for each SMG following \citet{Kennicutt98}:  
${\rm SFR}\,({\rm M}_{\odot}\,{\rm
  yr}^{-1})=1.7\times10^{-10}\,{\rm L}_{\rm IR}({\rm L_\odot})$.  This
relation assumes that the IR luminosity is predominantly powered by star formation
(i.e.~a negligible contribution from an AGN, which is a good
assumption in general for SMGs; e.g.~\citealt{Alexander05a,Alexander05b}; \citealt{Pope08};
\citealt{KMD10}; \citealt{Coppin10}; see also Section~\ref{spectra}) and comes from a starburst 
less than 100\,Myr old with a \citet{Salpeter55} IMF. This
relation also assumes that L$_{\rm IR}$ is calculated between
8--1000\,$\mu$m.  Because our single temperature modified
blackbody model does not reproduce the Wien side of the far-IR
SED needed in order to infer the full rest-frame L$_{\rm IR}$ from
8--1000\,$\mu$m (e.g.~\citealt{Blain03}), we use the approach of
\citet{Magnelli12} for calculating L$_{\rm IR}$: we have inferred  L$_{\rm IR}$
using the definition of \citet{Helou88}, L$_{\rm
  FIR}[42.5$--122.5\,$\mu$m], and applied a bolometric correction
factor of 1.91 from \citet{Dale01}.  As \citet{Magnelli12} note, this
introduces uncertainties in our L$_{\rm IR}$ estimates of the order of
$\pm30$ per cent due to the variation of intrinsic SED shapes, which
is not included in our quoted SFR uncertainties.

Table~\ref{tab:ir} lists our SED-fit derived FIR, L$_{\rm FIR}$,
T$_{\rm d}$ and SFR for each SMG, and Fig.~\ref{fig:seds} gives our
best-fitting SEDs.  The SED fitting has confirmed that our sample of SMGs
are far-IR luminous galaxies with L$_{\rm FIR} \simeq
0.6$--3\,$\times10^{12}$\,L$_{\odot}$ (and correspondingly L$_{\rm IR} \simeq
0.5$--5\,$\times10^{12}$\,L$_{\odot}$) and T$_{\rm d}\simeq
20$--40\,K and harbour intense
starbursts with SFRs of $\simeq 100$--1000\,M$_{\odot}$\,yr$^{-1}$,
consistent with (while lying towards the lower luminosity end of the
distribution) previous SMG SED measurements
(e.g.~\citealt{Kovacs06}; \citealt{Coppin08}; \citealt{Magnelli12}),
and reminiscent of the local (U)LIRG population \citep{SandMir96}.

\subsection{Probing the \hbox{[OI]}63\,$\mu$m strength in high-redshift SMGs}\label{ratio}

It has been an important objective of recent studies to measure the
strength of the major dense ISM cooling emission lines relative to the
total FIR emission.  Here we investigate this for the important
[OI]63\,$\mu$m $^{3}P_{2}\rightarrow^{3}P_{1}$ fine structure line,
which along with [OI]146\,$\mu$m and [CII]158\,$\mu$m, is a major
coolant in the transition layers between the molecular and atomic gas
in so called PDRs. The [OI]63\,$\mu$m emission line has been relatively unexplored at
high-redshift, but a well-known `deficit' or decrease in the strength of PDR cooling lines relative to the far-IR
continuum emission as a function of L$_{\rm FIR}$ has been observed in local
infrared-luminous galaxies (e.g.~\citealt{Luhman03};
\citealt{JGC11}).  Our new {\it Herschel}-PACS data can test whether this deficit extends from that seen in local
(U)LIRGs to high-redshift SMGs.

The SMGs detected in [OI]63\,$\mu$m emission in our sample have an [OI]63\,$\mu$m--to--FIR
ratio of $\simeq$0.5--1.5\,per cent.  Fig.~\ref{fig:main} shows the [OI]63\,$\mu$m emission line to FIR continuum ratio plotted
against L$_{\rm FIR}$ for the SMGs alongside all other low and high-redshift
galaxy populations with measured [OI]63\,$\mu$m emission available
from the literature.  The plot reveals that our SMGs do not
show an [OI]63\,$\mu$m deficit, which is different from what is seen in
local ULIRGs (galaxies with L$_{\rm FIR}\gsim10^{12}$\,L$_\odot$ in
Fig.~\ref{fig:main}), the purported local analogues of SMGs.  It thus appears
that [OI]63\,$\mu$m is a reliable tracer of star formation in high-redshift
SMGs. Our observations suggest that local ULIRGs and AGNs are not good analogues of typical
SMGs, and that SMGs appear to be more like local normal/starburst galaxies in terms of
their PDR cooling line properties relative to their far-IR emission,
which we now discuss in detail.

\section{Discussion}\label{discuss}

The [CII]158\,$\mu$m fine-structure emission line is the predominant coolant of the cold neutral
medium, since the C$^{+}$ ion is abundant and the 158\,$\mu$m emission
line is relatively easy to excite ($\Delta\,E/k \approx 92$\,K).
In warmer and denser environments (where $G_{\rm 0} \geq
10^{3}$ and $n \gtrsim 3\times10^{5}$\,cm$^{-3}$), 
[OI]63\,$\mu$m becomes the dominant coolant of the interstellar gas
($\Delta\,E/k \approx 228$\,K; \citealt{HT99}).  A
substantial fraction of the [OI]63\,$\mu$m emission is believed to originate in PDRs
at the interfaces between the H{\sc II} regions and molecular clouds
(e.g.~\citealt{Malhotra01}; \citealt{Contursi02}), with additional
contributions to the line emission originating from the diffuse
neutral medium or shocks (e.g.~\citealt{SpinoglioMalkan92}; \citealt{vanderWerf93}). 
[OI]63\,$\mu$m is thus an ideal
tracer of the warm, dense, neutral ISM, where star formation occurs.

The detection of [OI]63\,$\mu$m in two SMGs in our sample thus confirms the
presence of a warm dense neutral ISM in these systems, and indicates
that [OI]63\,$\mu$m is an important cooling line for SMGs.  The
marginal detection and the non-detections are also consistent with this
picture.  If [OI]63\,$\mu$m is the dominant cooling channel, then the
[OI]63\,$\mu$m--to--FIR ratio can be interpreted as a proxy of the gas heating efficiency
through the dust photo-electric effect for a relatively dense gas
($n\sim10^{5}$\,cm$^{-3}$; \citealt{Meijerink07};
\citealt{Kaufman99}).  If on the other hand [CII]158\,$\mu$m
contributes significantly to the cooling, then we can place a lower
limit to this heating efficiency.  Only such dense gas can excite the
[OI]63\,$\mu$m emission line and achieve a relatively high gas heating
efficiency of $\gsim1$\%, since the dense gas helps to keep the dust
grain charge low and thus the photo-electric yield high.

But what can these observations tell us about the star formation mode in SMGs, i.e.~how
their immense luminosities are powered? 
The [OI]63\,$\mu$m--to--FIR ratios seen in local starbursts are
$\simeq0.3$\%, and even higher ratios are seen in more `normal'
star formation driven galaxies (\citealt{Lord95}; \citealt{Malhotra01};
\citealt{JGC11}).  As mentioned above, local ULIRGs tend to show
an overall `deficit' in this ratio (and also for other important PDR emission
lines; e.g.~\citealt{JGC11}) relative to their less extreme
counterparts (see e.g.~Fig.~\ref{fig:main}).  One possible cause of this observed `deficit' is thought to be due to high values of the
ionization parameter (ie.~the number of hydrogen ionizing photons,
$h\,\nu > 13.6$\,eV, per hydrogen particle) at the surface of the
clouds (e.g.~\citealt{Abel05}; \citealt{Abel09}; \citealt{JGC11}). The effect of increasing the
ionization parameter causes a larger fraction of the UV photons to be
absorbed by dust (which are re-emitted in the IR) rather than being
available to ionize and excite the gas, resulting in a decreased
observed [OI]63\,$\mu$m--to--FIR ratio.  Thus, high ionization
parameters are the most likely cause of the PDR line deficit seen for
local (U)LIRGs \citep{JGC11}, which would indicate that there are
intrinsic differences in the mechanism or distribution of
star formation in these galaxies (ie.~more concentrated,
merger-induced) relative to more `normal' starburst galaxies
(ie.~more extended, quiescent). 

The fact that we do not see this deficit
in high-redshift luminous star formation dominated galaxies such as
SMGs suggests that the physics dominating the
[OI]63\,$\mu$m--to--FIR ratio must be similar to that dominating the star formation
activity in `normal' star-forming IR-bright galaxies, which have far-IR emission arising from an extended,
cooler `cirrus' component (e.g.~\citealt{ERR03}).  Importantly, the SMGs lie about an order of
magnitude above the [OI]63\,$\mu$m--to--FIR ratios of local ULIRGs and
AGNs, indicating that SMGs are not simple scaled up analogues of local ULIRGs, where the
star formation occurs in a compact and highly obscured nuclear region
(and potentially with a significant contribution from an AGN).  

Our result supports what is seen in the [CII]158\,$\mu$m emission line for
SMGs.  In a first modest-sized survey of the [CII]158\,$\mu$m emission
line in a heterogeneous sample of 13 far-IR luminous $1<z<2$ galaxies,
\citet{Stacey10} find that the luminous high-redshift star formation
dominated systems in their sample (including one gravitationally lensed and one unlensed SMG) do not show a
[CII]\,158\,$\mu$m deficit; whereas the AGN-dominated systems in their
sample do.  \citet{Stacey10} and \citet{DeBreuck11} use the observed
[CII]158\,$\mu$m--to--FIR ratio (which traces the far-UV field strength, $G_{\rm 0}$) and scaling arguments to constrain the
scale sizes of the three SMGs to $\simeq2$--5\,kpc \citep{Wolfire90}.  \citet{Ivison10} also report a [CII]158\,$\mu$m--to--FIR ratio
in SMM\,J2135--0102 (a lensed $z=2.3$ SMG) which is consistent with the SMG
being powered by starburst clumps distributed across $\sim2$\,kpc
\citep{Swinbank10}.  These studies add to the growing body of evidence
that high-redshift SMGs are host to extensive, galaxy-wide starbursts, in stark contrast to
the relatively more confined starbursts seen in local ULIRGs
(typically $\leq 100$\,s pc; e.g.~\citealt{Iono09}).  These
galaxy-wide scale sizes are similar to high resolution interferometric
mapping of the dense CO gas reservoirs of $1<z<4$ 
SMGs ($\simeq4$\,kpc;
\citealt{Tacconi06,Tacconi08}; \citealt{Bothwell10};
\citealt{Engel10}; \citealt{Hodge12}) and of the radio and
submm continuum and CO(1--0) emission 
($\simeq5$\,kpc up to 25\,kpc; \citealt{Chapman04};
\citealt{Younger08}; \citealt{Biggs08}; \citealt{Biggs10}; \citealt{Ivison11}).

Our new [OI]63\,$\mu$m measurements have provided a first step 
towards a census of the bright PDR
lines in SMGs.  We have shown that the SMGs in our sample do not show 
an [OI]63\,$\mu$m line deficit, similar to what was seen previously in
[CII]158\,$\mu$m, suggestive that this
is likely a global property for high-redshift luminous star-formation-dominated galaxies.  These SMGs are thus ideal candidates for future
observations of CO transitions and of the [CII]158\,$\mu$m emission line
-- which together could be used to provide a well-constrained PDR
solution for the emitting gas (see e.g.~\citealt{Stacey10}) to solve
for the density, $n$, and the far-UV radiation field strength, $G_{\rm
  0}$, in the ISM in SMGs.  Interestingly, it is predicted that [OI]63\,$\mu$m would
become more important than [CII]158\,$\mu$m as the galaxy
ISM heats up, which is most likely to occur during the most extreme
star formation events such as in SMGs.  At the earliest stages of
the starburst event, the C/O ratio is predicted to evolve strongly -- with
[OI]63\,$\mu$m predicted to be stronger than [CII]158\,$\mu$m emission
by at least a factor of $\sim3$ (e.g.~\citealt{Henkel93}; \citealt{Meijerink07};
\citealt{Kaufman07}), since [OI]63\,$\mu$m traces denser
gas ($n\sim3\times10^{5}$\,cm$^{-3}$) than [CII]158\,$\mu$m
($n\sim3\times10^{3}$\,cm$^{-3}$). Thus, in principle measuring the C/O ratio
could help to indicate the evolutionary state of the star formation
episodes of SMGs, using chemical evolutionary models such as those of 
\citet{Pipino04}. These models predict that throughout the starburst
episode, the C/O ratio would evolve strongly (since C enrichment
occurs mainly from low--to--intermediate mass stars) and thus this
ratio could be used as an effective way to `date' a galaxy and its
burst timescale in the early throes of star formation
(e.g.~\citealt{Maiolino05}).  Thus, a more complete set of diagnostic emission line data would
begin to reveal a fuller picture of early nucleosynthesis and
interstellar processes in high-redshift luminous star-forming
galaxies, which is critical to our understanding of the driving
mechanisms involved in these modes of star formation.  

\section{Conclusions}\label{conclusions}

We used {\it Herschel}-PACS to target the [OI]63\,$\mu$m
emission line in a sample of {\it unlensed} high-redshift
870\,$\mu$m-selected $1.1<z<1.6$ SMGs for the first
time.  Our sample originally consisted of nine statistically robust
radio and/or mid-IR counterparts to the SMGs (providing precise
positions for the SMGs) with high-quality
spectroscopic redshifts.  We used high-resolution interferometric ALMA
870\,$\mu$m continuum data to verify the reliability of our original target
selection.  The ALMA data have verified unambiguously that six
of the SMG counterparts are bona fide
submm-emitting sources; three of the SMG
counterparts that were targeted here do not appear to be a source of submm emission and we have
removed these from our sample. 

We detect the [OI]63\,$\mu$m emission line in two of the
SMGs, tentatively detect [OI]63\,$\mu$m emission in one SMG, and measure sensitive 3\,$\sigma$ upper limits to the line
flux in the remaining three cases, secure in the knowledge that we
have targeted the source of the submm emission and the correct
redshift.  The [OI]63\,$\mu$m detections confirm the presence of a warm dense neutral ISM in these
systems and indicate that [OI]63\,$\mu$m is an important PDR cooling line for SMGs.
We find that the [OI]63\,$\mu$m--to--FIR ratio in SMGs (ranging from
$\simeq$0.5--1.5\,per cent) does not follow the well-documented PDR
cooling line `deficit' that has been observed for local ULIRGs and
some local AGNs which are in a similar far-IR luminosity class to SMGs. This result suggests
that local ULIRGs and AGNs are not good analogues of typical SMGs, and
that SMGs appear to be more like `normal' starburst galaxies in terms of
their star formation properties.  This offset in the [OI]63\,$\mu$m--to--FIR
ratio compared to local (U)LIRGs has been noted previously for a small sample of SMGs in another
classical PDR emission line, [CII]158\,$\mu$m, and together these
results suggest that this could be a global property for all the
bright PDR cooling lines for SMGs. We have taken advantage of a unique window of opportunity
provided by {\it Herschel} to verify that [OI]63\,$\mu$m is thus a reliable tracer of star formation in high-redshift
star formation dominated galaxies such as SMGs.  These \textit{Herschel} observations are 
a necessary stepping stone to future ALMA studies of [CII]158\,$\mu$m
and CO emission in order to investigate the ISM physics
of SMGs in detail. 

\section*{Acknowledgments}

We thank an anonymous referee for a helpful report which improved the
clarity of the paper. 
We are sincerely grateful to Eckhard Sturm and Javier Graci\'{a}-Carpio for advice
and analysis tips for the reduction of {\it Herschel} spectroscopy data for
faint high-redshift lines and for kindly providing their literature
compilation data points.  We would also like to thank the staff at the
NASA {\it Herschel} Science Centre for hosting very informative and
useful data reduction workshops which played a key role in the success and
completion of this programme.  KEKC acknowledges support from the endowment of the Lorne Trottier
Chair in Astrophysics and Cosmology at McGill, the Natural Sciences and Engineering
Research Council of Canada (NSERC), and a L'Or\'{e}al Canada for Women in Science
Research Excellence Fellowship, with the support of the Canadian
Commission for UNESCO. WNB acknowledges the support of a NASA {\it Herschel} grant RSA
1438954 and a NASA ADP grant NNX10AC99G. JSD acknowledges the support of the European Research Council via
the award of an Advanced Grant, and the support of the Royal Society via a Wolfson Research Merit Award.
JEG is supported by an NSERC Banting Postdoctoral Fellowship. TRG
acknowledges support from the UK Science and Technologies Facilities
Council for support, as well as IDA and DARK. 
The {\it Herschel} spacecraft was designed, built, tested, and
launched under a contract to ESA managed by the {\it Herschel/Planck} Project team by an industrial consortium under the overall responsibility of the prime contractor Thales Alenia Space (Cannes), and including Astrium (Friedrichshafen) responsible for the payload module and for system testing at spacecraft level, Thales Alenia Space (Turin) responsible for the service module, and Astrium (Toulouse) responsible for the telescope, with in excess of a hundred subcontractors.
This research has made use of {\sc hipe}, which is a joint development
by the {\it Herschel} Science Ground Segment Consortium, consisting of
ESA, the NASA {\it Herschel} Science Center, and the HIFI, PACS and SPIRE consortia.

\setlength{\bibhang}{2.0em}

\end{document}